\documentclass[a4paper,11pt]{article}
\pdfoutput=1 

\usepackage{jheppub} 

\usepackage[T1]{fontenc} 
\usepackage{slashed}
\usepackage{tikz}
\usetikzlibrary{shapes,arrows,positioning,automata,backgrounds,calc,er,patterns}
\usepackage{tikz-feynman}
\tikzfeynmanset{compat=1.1.0}
\usepackage{xcolor}
\usepackage{makecell}
\usepackage{rotating}
\usepackage{pdflscape}
\usepackage{float}
\newcommand{\Tr}[4]{\mathrm{Tr}(\slashed{#1}\slashed{#2} \slashed{#3} \slashed{#4})}
\newcommand{\ome}{(1-\epsilon)}
\newcommand{\omxi}{(1-x_i)}

\newcommand{\omxk}{(1-x_k)}
\newcommand{\e}{\epsilon}

\newcommand{\fa}{a}
\newcommand{\fb}{b}
\newcommand{\fc}{\tilde{b}}
\newcommand{\fd}{c}
\newcommand{\fe}{\tilde{c}}
\newcommand{\ff}{d}
\newcommand{\fg}{\tilde{d}}

\newcommand{\Pqg}{P_{qg}}
\newcommand{\Pgq}{P_{gq}}
\newcommand{\Pgg}{P_{gg}}
\newcommand{\PggS}{P_{gg}^\text{sub}}
\newcommand{\Pqq}{P_{q\bar{q}}}
\newcommand\nameggg{ggg \to g}

\newcommand\nameqgg{qgg \to q}
\newcommand\nameqpp{q\gamma\gamma \to q}
\newcommand\namegqbq{g\bar q q\to g}
\newcommand\namepqbq{q g \bar q \to \gamma}
\newcommand\nameqQQ{q\bar Q Q\to q}
\newcommand\nameqqq{q \bar q q\to q}
\newcommand{\Pggg}{P_{\nameggg}}
\newcommand{\Rggg}{R_{\nameggg}}

\newcommand{\Rgggsub}{R_{g(gg)}}
\newcommand{\Pqgg}{P_{\nameqgg}}
\newcommand{\Rqgg}{R_{\nameqgg}}
\newcommand{\Pqpp}{P_{\nameqpp}}
\newcommand{\Rqpp}{R_{\nameqpp}}
\newcommand{\Pgqbq}{P_{\namegqbq}}
\newcommand{\Rgqbq}{R_{\namegqbq}}
\newcommand{\Ppqbq}{P_{\namepqbq}}
\newcommand{\Rpqbq}{R_{\namepqbq}}
\newcommand{\PqQQ}{P_{\nameqQQ}}
\newcommand{\RqQQ}{R_{\nameqQQ}}
\newcommand{\Pqqq}{P_{\nameqqq}}
\newcommand{\Rqqq}{R_{\nameqqq}}

\title{\boldmath Decomposition of Triple Collinear Splitting Functions}

\author{Oscar Braun-White}
\author{and Nigel Glover}

\affiliation{Institute for Particle Physics Phenomenology,\\Department of Physics, \\Durham University, Durham, DH1 3LE, UK}

\emailAdd{oscar.r.braun-white@durham.ac.uk}
\emailAdd{e.w.n.glover@durham.ac.uk}

\preprint{IPPP/22/21}

\abstract{In the kinematic region where three particles $i$, $j$, $k$ are collinear, the multi-parton scattering amplitudes factorise into a product of a triple collinear splitting function and a multi-parton scattering amplitude with two fewer particles. These triple collinear splitting functions contain both iterated single unresolved contributions, and genuine double unresolved contributions. We make this explicit by rewriting the known triple collinear splitting functions in terms of products of two-particle splitting functions, and a remainder that is explicitly finite when any two of $\{i,j,k\}$ are collinear. We analyse all of the single unresolved singularities present in the remainder.
}

\begin{document}

\maketitle
\flushbottom

\section{Introduction \label{sec:intro}}

The experiments at the Large Hadron Collider (LHC) are able to measure particle scattering with unprecedented precision, approaching the percent level for some observables.  Precise theoretical predictions that are adapted to the specific experimental observables and that match the accuracy of the experimental measurements are needed to extract fundamental Standard Model parameters.  Typically, theoretical predictions are obtained using perturbation theory as an expansion in the coupling.  The precision of the theoretical predictions is generally limited by a dependence on unphysical renormalisation and factorisation scales, or through the modelling of complicated final states with relatively few final state particles. This can be systematically improved by including higher-order corrections. The leading order (LO) prediction captures the gross features of an observable. Inclusion of next-to-leading order (NLO) corrections is required to estimate the normalisation of the predictions. Even higher orders (NNLO, N3LO, \ldots) are needed to describe detailed event properties or to achieve the goal of percent level precision.

In general, there are two obstacles to the perturbative expansion.  First, knowledge of the relevant tree and loop multiparticle scattering amplitudes. In the framework of dimensional regularisation, gauge-theory loop amplitudes contain explicit infrared poles in the regulator $\e$ of up to two powers per loop. The computation of such amplitudes is sufficiently complicated that it is a field in its own right. Second, a scheme to extract the implicit infrared divergences.   These are produced by integration of amplitudes with fewer loops and more external particles over the unresolved or infrared-singular regions of the phase space. The explicit poles and implicit poles are cancelled in physical cross sections, thereby enabling the numerical evaluation over the whole of phase space.  

For multiparticle final states (corresponding to $2 \to 4,5,\ldots$ kinematics), the state-of-the-art are NLO perturbative corrections.  Automated programmes exist for calculating tree and one-loop amplitudes together with the necessary infrared subtraction terms. These are encapsulated in a number of multi-purpose event generator programs~\cite{herwig:2015jjp,Sherpa:2019gpd,powheg:2010xd,madgraph:2011uj}, enabling NLO-accurate predictions for essentially any relevant collider process.\footnote{See Ref.~\cite{snowmass:2022qmc} for a summary of available tools.} At NNLO, calculations are mostly limited to $2\to 2$ kinematics like, for example, di-jet production~\cite{Currie:2016bfm,Czakon:2019tmo}, vector-boson-plus-jet production~\cite{Boughezal:2015dva,Gehrmann-DeRidder:2015wbt, Boughezal:2015ded}, photon-plus-jet-production~\cite{Campbell:2016lzl,Chen:2019zmr} or top quark pair production~\cite{Czakon:2015owf,Catani:2019iny}.  Recent progress in the derivation of two-loop $2\to 3$ scattering amplitudes has led to calculations for three-photon production~\cite{Chawdhry:2019bji}, diphoton-plus-jet production~\cite{Chawdhry:2021hkp} and three-jet production~\cite{Czakon:2021mjy}. Several infrared subtraction methods have been developed for NNLO calculations (See Ref.~\cite{TorresBobadilla:2020ekr} for a review.). Implementations using these methods are largely made on a process-by-process basis, and most methods scale either poorly or not at all to higher multiplicities. 

Recently, the first steps towards N3LO calculations for $2 \to 1$ processes have been taken, with the computations of fully inclusive coefficient functions for Higgs production~\cite{Anastasiou:2016cez,Dulat:2018bfe} and the Drell-Yan process~\cite{Duhr:2020seh,Chen:2021isd}, which are now being extended towards fully differential final states~\cite{Chen:2021vtu,Billis:2021ecs,Chen:2022cgv,Neumann:2022lft}. The infrared subtraction methods used for N3LO calculations exploit the very special $2 \to 1$ kinematics, and no systematic method has been established.

The need for NNLO and N3LO predictions for phenomenologically relevant high-multiplicity processes highlights the importance of developing a more systematic and structured infrared subtraction formalism.  

The universal factorisation properties of multiparticle amplitudes are important for generating counter terms that can be used to isolate the infrared singularities that are produced in particular regions of phase space, when one or more particles are unresolved.\footnote{See Ref.~\cite{Agarwal:2021ais} for a review. These factorisation properties are also key in quantifying the accuracy of parton branching algorithms in event generators, and how these algorithms can eventually be extended to increase their logarithmic accuracy, see for example Refs.~\cite{Li:2016yez,Hoche:2017iem,Dulat:2018bfe,Dulat:2018vuy,Dasgupta:2020fwr,Loschner:2021keu,Gellersen:2021eci}} Most well studied are the single unresolved limits, where either one particle is soft, or two are collinear, which are relevant for NLO calculations.  At NNLO, one is concerned with the double unresolved limits of tree amplitudes~\cite{campbell,Catani:1998nv,Catani:1999ss,Kosower:2002su}, as well as the single unresolved limit of one-loop amplitudes~\cite{Bern:1994zx,Bern:1998sc,Kosower:1999rx,Bern:1999ry}.  
At N3LO, one encounters the triple unresolved limits of tree amplitudes~\cite{Catani:2019nqv,DelDuca:1999iql,DelDuca:2019ggv,DelDuca:2020vst,DelDuca:2022noh}, the double unresolved limits of one-loop amplitudes~\cite{Catani:2003vu,Sborlini:2014mpa,Badger:2015cxa,Zhu:2020ftr,Catani:2021kcy,Czakon:2022fqi} and the single unresolved limits of two-loop amplitudes~\cite{Bern:2004cz,Badger:2004uk,Duhr:2014nda,Li:2013lsa,Duhr:2013msa}.  And so on.

One of the complications immediately evident at NNLO is the overlap between iterated single unresolved and genuinely double unresolved limits. For example, the limit in which three particles $i$, $j$ and $k$ become collinear (studied in Refs.~\cite{campbell,Catani:1998nv,Catani:1999ss}) is obtained when invariants in the set $\{s_{ij}$, $s_{jk}$, $s_{ik}$, $s_{ijk} \}$ are small and there are two inverse powers of them. This limit contains both single and double unresolved limits - an iterated collinear contribution (which overlaps with soft and collinear limits), as well as a genuinely double unresolved contribution.  In this paper we decompose the triple collinear limits into products of two-particle splitting functions, and a remainder that is explicitly finite when any two of $\{i,j,k\}$ are collinear.
 
To help with the discussion of the singularities present in the real radiation amplitudes, we introduce the notion of {\bf internal} and {\bf external} singularities. Internal singularities are associated with small invariants amongst the set of collinear particles. External singularities involve other (spectator) particles involved in the scattering through the definition of the momentum fraction. For example, when two particles, $i$ and $j$ are collinear we find the well known single collinear limit proportional to the two-particle splitting function, 
$$
\frac{1}{s_{ij}} P_{ab}(x_i).
$$
The limit as $s_{ij} \to 0$ references only particles in the collinear set and is therefore an internal singularity. External singularities are both present in the splitting functon $P_{ab}(x_i)$ and associated with the momentum fraction limits $x_i \to 0$ or $x_i \to 1$. These external singularities correspond to situations where one of $\{i,j\}$ is collinear with a spectator particle, or where one of the particles is soft.

As in Ref.~\cite{campbell}, we work with colour-ordered amplitudes and consider spin-averaged collinear limits, which are directly obtained by taking the collinear limit of partonic ``squared'' matrix elements. One could equivalently work in colour space, and retain information about the spin of the parton formed from the merger of the collinear particles, as was done in Refs.~~\cite{Catani:1998nv,Catani:1999ss}. The spin-unaveraged splitting functions contain additional azimuthal correlations when the parent parton is a gluon that reflect different orientations of the final state particles with respect to the gluon polarisation (and effectively with respect to other particles not involved in the triple collinear limit).  These azimuthal correlations are not present in the case where the parent parton is a quark, since the splitting function is proportional to the unit matrix in the spin indices. 

Our paper is organised as follows.  Section~\ref{sec:notation} establishes our notation. We discuss the infrared singularity structure of the triple collinear splitting functions in Section~\ref{sec:singstructure}. In Section~\ref{sec:genstructure} we discuss the general structure of the triple collinear limit, and explain how to restructure it such that the strongly-ordered limit is explicit, and the remaining terms are manifestly finite when any two of $\{i,j,k\}$ are collinear. 
Results for the triple collinear splitting function for all of the various parton configurations are collected in Section~\ref{sec:results}.  We also analyse all of the internal and external single unresolved singularities of each of the splitting functions. Finally, we summarise our findings in Section~\ref{sec:summary}.

\section{Notation \label{sec:notation}}

We consider the time-like triple collinear limits of colour-connected particles that were first discussed in Ref.~\cite{campbell}. Following the notation of Ref.~\cite{campbell}, we employ colour-ordered amplitudes. If particle $i$ is colour-connected to $j$ which is colour-connected to $k$, the colour-ordered amplitude is given by,
\begin{equation}
    \mathcal{A} (...,i,j,k,...).
\end{equation}
In the limit that three colour-connected particles become collinear, the squared colour-ordered amplitude factorises as, 
\begin{equation}
\label{eq:TC1}
	|\mathcal{A} (...,i,j,k,...)|^2 \rightarrow P_{abc \to P}(i,j,k) |\mathcal{A} (...,P,...)|^2 .
\end{equation}
Here,  $i,j,k$ are labels for three colour-connected partons of particle type $a,b,c$ with four-momenta $p_i^{\mu}$, $p_j^{\mu}$ and $p_k^{\mu}$ which become collinear in a process involving four or more partons. 
In Eq.~\eqref{eq:TC1}, $i$, $j$ and $k$ are all colour-connected.  
There are also configurations in which particles that are not colour-connected can usefully be thought of as colour-connected. This happens when there is more than one colour-string - there is an antiquark at the end of one colour-string and a like flavour quark at the beginning of another. For example, the amplitude
\begin{equation}
    \mathcal{A} (...,\bar{Q}|Q,...)
\end{equation}
represents a situation where there are two colour-strings, one terminated by the fundamental colour index of the $\bar{Q}$ and another initiated by the fundamental colour index of the $Q$. In this case, when the quark-antiquark pair are collinear, they combine to form a gluon, which then connects, or pinches together, the two colour-strings,
\begin{equation}
\label{eq:TC2}
	|\mathcal{A} (...,i,j|k...)|^2 \rightarrow P_{a\bar{Q}Q \to P}(i,j,k) |\mathcal{A} (..,P,...)|^2 .
\end{equation}

In the triple collinear limit, the collinear cluster has momentum 
\begin{equation*}
p_i^{\mu} + p_j^{\mu}+ p_k^{\mu} = p_P^\mu.
\end{equation*}
We define Lorentz invariant quantities,
\begin{equation}
	s_{i,\ldots,n} \equiv (p_{i}+...+p_{n})^2.
\end{equation}
For massless quarks and gluons, $s_{ij} = 2p_i \cdot p_j = 2E_iE_j(1-\cos\theta_{ij})$, where $E_i$, $E_j$ are the energies of particles $i$, $j$ and $\theta_{ij}$ is the angle between them. $s_{ij}$ approaches zero if either particle is soft or they are collinear. We systematically work in dimensional regularisation with $d=4-2\epsilon$. 
The triple collinear limit is defined as the kinematic regime where the invariants $s_{ij},$ $s_{jk}$, $s_{ik}$, $s_{ijk}$ become small and therefore $p_P^2 \sim 0$. In this limit, we can write $p_i = x_i p_P$, $p_j = x_j p_P$ and $p_k = x_k p_P$ with $x_i+x_j+x_k = 1$.  In practice, a spectator momentum $\ell$ is used to define the momentum fractions, $s_{i\ell}= x_i s_{P\ell}$.

The particle $P$ retains the quantum numbers of the collinear partons and there are seven possible clusterings: $\nameggg$, $\nameqgg$, $\nameqpp$, $\namegqbq$, $\namepqbq$, $\nameqQQ$ and $\nameqqq$. 
The triple collinear splitting functions depend on the momentum fractions and the small invariants.  However, for brevity we will suppress these arguments and use a shorthand notation,
\begin{equation}
P_{abc\to P} (i,j,k) \equiv P_{abc \to P}(x_i, x_j, x_k; s_{ij},s_{ik},s_{jk},s_{ijk}).
\end{equation} 

\section{Singularity structure of the triple collinear splitting function} \label{sec:singstructure}

The primary aim of this paper is to rewrite the $P_{abc \to P}$ splitting function in a way that exposes its singularity structure. In particular, we aim to isolate the strongly-ordered iterated contributions.  In other words, we aim to rewrite the spin-averaged and colour-ordered three-particle splitting function as,
\begin{equation}
\label{eq:Psplit}
P_{abc\to P} (i,j,k) = \sum_{{\mathrm perms }} 
\frac{1}{s_{ijk}} P_{(ab)c \to P}\left(x_k\right)
\frac{1}{s_{ij}} P_{ab \to (ab)}\left(\frac{x_j}{1-x_k}\right) 
+ \frac{1}{s_{ijk}^2}R_{abc\to P} (i,j,k)
\end{equation}
where $P_{ab \to (ab)}$ are the usual spin-averaged two-particle splitting functions (listed in the Appendix) and the remainder $R_{abc\to P} (i,j,k)$ depends on the momentum fractions and small invariants. 

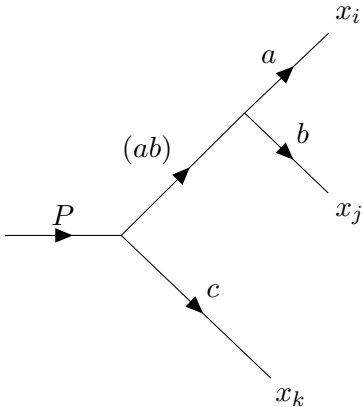
\begin{figure}[th!]
    \centering

\begin{tikzpicture} 
\begin{feynman}
\vertex (a); 
\vertex [right=4em of a] (b);
\vertex [above right=6em of b]  (d) ; 
\vertex [above right=of d]      (f1) {\(x_i\)};
\vertex [below right=of d]      (f2) {\(x_j\)};
\vertex [below right=7em of b]  (f3) {\(x_k\)};

\diagram* {
(a) -- [fermion, edge label=\( P \)] (b),
(b) -- [fermion, edge label=\( (ab) \)] (d), 
(d) -- [fermion, edge label=\( a \)] (f1), 
(d) -- [fermion, edge label=\( b \)] (f2),
(b) -- [fermion, edge label=\( c \)] (f3),
};
\end{feynman}
\end{tikzpicture}

    \caption{The iterated single-collinear contribution to the triple collinear splitting function.  }
    \label{fig:iterated}
\end{figure}

\begin{figure}[th!]
    \centering
\begin{tikzpicture} 
\begin{feynman}
\vertex (a); 
\vertex [right=4em of a] (b);
\vertex [above right=7em of b]  (f1) {\(x_i\)};
\vertex [right=6em of b]        (f2) {\(x_j\)};
\vertex [below right=7em of b]  (f3) {\(x_k\)};

\diagram* {
(a) -- [fermion, edge label=\( P \) ] (b),
(b) -- [fermion, edge label=\( a \)] (f1),
(b) -- [fermion, edge label=\( b \)] (f2),
(b) -- [fermion, edge label=\( c \)] (f3), 
};
\end{feynman}
\end{tikzpicture}            

    \caption{The remainder function $R_{abc\to P}$ contains the parts of the triple collinear splitting function that are not contained in the strongly-ordered, iterated contribution.}
    \label{fig:Rijk}
\end{figure}
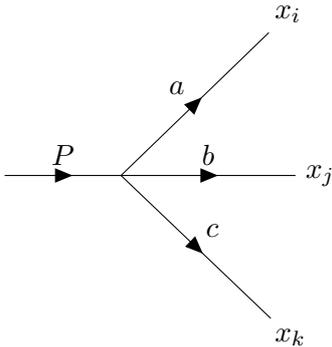

An iterated (or strongly-ordered) contribution is obtained through the product of leading-order splitting functions, $P \times P$, as illustrated in ~Fig.~\ref{fig:iterated} and is given by terms of the type,
\begin{equation}
    \frac{P_{(ab)c}(x_k)}{s_{ijk}} \times
     \frac{P_{ab}(y_j)}{s_{ij}} 
\end{equation}  
where $y_j$ is the momentum fraction of the second splitting, 
$$
y_j = \frac{x_j}{x_i+x_j} = \frac{x_j}{1-x_k}.
$$
The invariants in the denominator are simply those corresponding to the two- and three-particle invariants, $s_{ij}$ and $s_{ijk}$.\footnote{Note that one could have chosen to define the strongly-ordered limit in which $s_{ijk}$ is replaced by $s_{ik}+s_{jk}$.} The remainder (or uniterated) $1\rightarrow 3$ splitting function $R_{abc\to P}$ is illustrated in Fig.~\ref{fig:Rijk}. 

The triple collinear splitting functions contain both single and double unresolved limits:
\begin{enumerate}
\item single collinear limits when two of $\{i,j,k\}$ are collinear
or one of $\{i,j,k\}$ is collinear with a spectator particle, 
\item single soft limits when one particle is soft and is colour-connected to either the other two particles, or a spectator particle,
\item double collinear limits when two of $\{i,j,k\}$ are collinear and the third is collinear with a spectator particle,
\item soft-collinear limits when one particle is soft and the other two are collinear or one is collinear with a spectator particle,
\item double soft limits when two particles are soft,
\item triple collinear limits when two of $\{i,j,k\}$ are collinear with a spectator particle (this occurs only in the double soft limits).
 
\end{enumerate}

In order to explain what types of single unresolved singularities appear in the iterated contributions, $P\times P$, and the uniterated splitting function, $R_{abc\to P}$, we must first discuss the different types of single unresolved singularities present in $P_{abc \to P}$. 
 
There are two types of single collinear singularities present in a triple collinear splitting function. First, internal single collinear singularities like $1/s_{ij}$, where $i,j$ are collinear. Internal single collinear singularities appear only in the iterated two-particle splitting contributions. Second, external single collinear singularities like $1/x_i$, which indicate potential collinear factors with the spectator particles used to define the momentum fractions. External single collinear singularities which are present in $P_{abc \to P}$ appear only in the iterated two-particle splitting contributions. Although, if $P_{abc \to P}$ does not contain external single collinear singularities, there could be a cancellation between external single collinear singularities in the $P\times P$ contribution and those in the remainder $R_{abc \to P}$. 

There are also two types of single soft singularities. First, internal soft $j$ singularities encoded through typical eikonal factors like $s_{ik}/(s_{ij} s_{jk})$. This type of singularity is explicitly embedded in a triple collinear splitting function and only appears in the remainder $R_{abc \to P}$. This makes sense because internal single soft singularities are inherently uniterated - this type of eikonal factor contributes the full weight of a triple collinear term without an $s_{ijk}$ pole. Second, there are external soft $j$ singularities that appear in `hidden' eikonal factors like $x_i/(x_j s_{ij})$. This type of singularity is produced when the soft particle is colour-connected to a spectator particle. If present, it only appears in the iterated two-particle splitting contributions.  

We can summarise these important features as follows:
\begin{itemize}
    \item[-] {\bf Internal} single collinear singularities like $1/s_{ij}$ appear only in $P \times P$ terms (the iterated two-particle splitting contributions). 
    \item[-] When {\bf external} single collinear singularities like $1/x_i$ appear in $P_{abc \to P}$, they are all contained in $P \times P$ terms.
    \item[-] When {\bf external} single collinear singularities like $1/x_i$ {\bf do not} appear in $P_{abc \to P}$, there could be terms proportional to $1/x_i$ in $P \times P$ and $R_{abc \to P}$ which cancel. 
    \item[-] {\bf Internal} single soft singularities like $s_{ik}/(s_{ij} s_{jk})$ appear only in $R_{abc \to P}$.
    \item[-] {\bf External} single soft singularities like $x_i/(x_j s_{ij})$ appear only in the iterated  $P \times P$ terms.
    \end{itemize}

\section{General structure of the triple collinear splitting function} \label{sec:genstructure}

As mentioned earlier, the triple collinear limit is defined as the kinematic regime where the invariants $s_{ij},$ $s_{jk}$, $s_{ik}$ and $s_{ijk}$ all become small. In this region, the singular factor has at most two net inverse powers of the small invariants. Additionally, since the splitting functions are limits of squared amplitudes, there is the additional physics constraint that there are at most two inverse powers of double invariants ($s_{IJ}$ where $I,J = \{i,j,k\}$) and at most two inverse powers of the triple invariant $s_{ijk}$. Therefore, any triple collinear splitting function can be represented by coefficients $\beta_i (x_i,x_j,x_k,\epsilon)$ of 37 invariant pole structures: 
\begin{equation} \label{eqn:spoles}
	\begin{split} 
P_{abc\to P}(i,j,k) =	& \color{black} \frac{\beta_1}{s_{jk}s_{ijk}} + \frac{\beta_2}{s_{ij}s_{ijk}}+ \frac{\beta_3}{s_{ik}s_{ijk}}  \color{black} + \frac{\beta_4}{s_{jk}^2}+ \frac{\beta_5}{s_{ij}^2}+ \frac{\beta_6}{s_{ik}^2} \\
	\color{black} +& \color{black} \frac{\beta_7}{s_{jk} s_{ij}} + \frac{\beta_8}{s_{jk}s_{ik}}+ \frac{\beta_9}{s_{ij}s_{ik}} + \frac{\beta_{10} s_{ij}}{s_{jk}^2 s_{ijk}} + \frac{\beta_{11} s_{ij}}{s_{ik}^2 s_{ijk}} + \frac{\beta_{12} s_{jk}}{s_{ij}^2 s_{ijk}}  \\
	\color{black} + & \color{black} \frac{\beta_{13}}{s_{ijk}^2}+  \frac{\beta_{14} s_{ij}}{s_{jk} s_{ijk}^2} + \frac{\beta_{15} s_{ij}}{s_{ik} s_{ijk}^2} + \frac{\beta_{16} s_{jk}}{s_{ij} s_{ijk}^2} 
	+ \frac{\beta_{17} s_{ij}^2}{s_{jk}^2 s_{ijk}^2}+ \frac{\beta_{18} s_{ij}^2}{s_{ik}^2 s_{ijk}^2} + \frac{\beta_{19} s_{jk}^2}{s_{ij}^2 s_{ijk}^2}\\
	\color{black} + & \color{black} \bigg[ \frac{\beta_{20} s_{jk}}{s_{ik}^2 s_{ijk}} + \frac{\beta_{21} s_{jk}}{s_{ik} s_{ijk}^2} + \frac{\beta_{22} s_{jk}^2}{s_{ik}^2 s_{ijk}^2} +  \frac{\beta_{23} s_{ij}^2}{s_{ik} s_{jk} s_{ijk}^2} 
	+ \frac{\beta_{24} s_{ij}}{s_{ik} s_{jk} s_{ijk}} + \frac{\beta_{25} s_{jk}}{s_{ik} s_{jk} s_{ijk}} \\
    \color{black} + & \color{black} \frac{\beta_{26} s_{ik}}{s_{ij}^2 s_{ijk}} +  \frac{\beta_{27} s_{ik}}{s_{jk}^2 s_{ijk}} +  \frac{\beta_{28} s_{ik}}{s_{ij} s_{ijk}^2} + \frac{\beta_{29} s_{ik}}{s_{jk} s_{ijk}^2} + \frac{\beta_{30} s_{ik}^2}{s_{ij}^2 s_{ijk}^2} + \frac{\beta_{31} s_{ik}^2}{s_{jk}^2 s_{ijk}^2} + \frac{\beta_{32} s_{ik}}{s_{ij} s_{jk} s_{ijk}} \\
    \color{black} +& \color{black} \frac{\beta_{33} s_{ik}^2}{s_{ij} s_{jk} s_{ijk}^2}  
    +\frac{\beta_{34} s_{jk}^2}{s_{ij} s_{ik} s_{ijk}^2}  + \frac{\beta_{35} s_{ij} s_{ik}}{s_{jk}^2 s_{ijk}^2} + \frac{\beta_{36} s_{ij} s_{jk}}{s_{ik}^2 s_{ijk}^2}+ \frac{\beta_{37} s_{ik} s_{jk}}{s_{ij}^2 s_{ijk}^2} \bigg].
\end{split}
\end{equation}
Using momentum conservation, any triple collinear splitting function can be expressed in the basis of the first three lines of Eq. (\ref{eqn:spoles}) (ie. $\beta_1$ --- $\beta_{19}$, the non-square bracketed terms). 

The factorisation properties of squared amplitudes impose six additional relationships amongst the $\beta_i$. Of these there are two relationships between $\beta_4$, $\beta_{10}$, $\beta_{17}$, which are due to the absence of $1/s_{jk}^2$ single collinear contributions. Similar relationships hold for the coefficients of $1/s_{ij}^2$ and $1/s_{ik}^2$. Therefore, we propose an alternative basis in terms of 13 $\alpha_i (x_i,x_j,x_k,\epsilon)$ invariant structures that make the physical constraints more evident,
\begin{eqnarray}
\label{eq:abasis}
P_{abc\to P}(i,j,k) &= &\phantom{+}
\frac{\alpha_{12}}{s_{jk}s_{ijk}}+ 
\frac{\alpha_{13}}{s_{ij}s_{ijk}}+
\frac{\alpha_{14}}{s_{ik}s_{ijk}} \nonumber \\
& & + \frac{\alpha_1}{s_{ijk}^2} 
+ \frac{\alpha_2 \Tr{j}{k}{i}{\ell}}{ s_{jk} s_{ijk}^2} 
+ \frac{\alpha_3 \Tr{i}{j}{k}{\ell}}{ s_{ij} s_{ijk}^2} 
+ \frac{\alpha_4 \Tr{k}{i}{j}{\ell}}{ s_{ik} s_{ijk}^2} \nonumber \\
& &
+ \frac{\alpha_{23} \Tr{i}{j}{k}{\ell}}{ s_{ij}  s_{jk} s_{ijk} } 
+ \frac{\alpha_{24} \Tr{j}{k}{i}{\ell}}{ s_{jk}  s_{ik} s_{ijk}} 
+ \frac{\alpha_{34} \Tr{k}{i}{j}{\ell}}{ s_{ik}  s_{ij} s_{ijk}}  \nonumber \\
& &
+ \frac{\alpha_{22} W_{jk}}{ s_{jk}^2 s_{ijk}^2} 
+ \frac{\alpha_{33} W_{ij}}{ s_{ij}^2 s_{ijk}^2} 
+ \frac{\alpha_{44} W_{ik}}{ s_{ik}^2 s_{ijk}^2} .
\end{eqnarray}
Here $\ell$ is a suitably normalised spectator momentum such that,
\begin{equation}
\Tr{i}{j}{k}{\ell} = x_k s_{ij} -x_j s_{ik} + x_i s_{jk} ,    
\end{equation}
while the quantity $W_{ij}$ is defined as
\begin{equation}
\label{eq:Wdef}
        W_{ij} = (x_i s_{jk} - x_j s_{ik})^2 - 
        \frac{ 2} {\ome}
        \frac{ x_i x_j x_k} {(1-x_k)}
        s_{ij} s_{ijk}.
\end{equation}
In Eq.~\eqref{eq:abasis}, the first three coefficients ($\alpha_{12}$, $\alpha_{13}$, $\alpha_{14}$) display any strongly-ordered structure present, like in Eq.~(\ref{eq:Psplit}).  
The remaining structures are written in combinations that are designed to be less singular in the single collinear limits. 
For example, in the $s_{ij} \to 0$ limit,
\begin{equation} \label{eq:Trsingularity}
\Tr{i}{j}{k}{\ell} = \mathcal{O} (\sqrt{s_{ij}})
\end{equation}
so that there is no singular contribution in the $ij$ collinear limit from the $\alpha_3$, $\alpha_{23}$ or $\alpha_{34}$ terms.\footnote{Note that an alternative basis to the $\alpha$ basis could be chosen with somewhat different structures to the trace structure used here. We choose the trace structure for its 'natural' interpretation and see that it reflects the colour-ordering in the results. Another suitable basis would require properties which follow equations similar to Eq.~\eqref{eq:Trsingularity}.}

Similarly, the $\alpha_{33}$ term also has no contribution in the $s_{ij} \rightarrow 0$ limit. $W_{ij}$ has been constructed from terms that appear in the triple collinear limit, and a second term that is added to $\alpha_{13}$ (and subtracted from $\alpha_{33}$ in order to have the full spin-averaged splitting functions in the strongly-ordered contributions).  Both terms in $W_{ij}$ are individually $\mathcal{O} (s_{ij})$ when expanded but have opposite signs so that $W_{ij}/s_{ij}^2 = \mathcal{O} (1/\sqrt{s_{ij}})$. This is an integrable singularity that vanishes upon azimuthal integration (in $d$-dimensions). To make this clear, strictly in the collinear $i$, $j$ limit, we can interpret $W_{ij}$ in terms of the azimuthal angle with respect to the $(ij)$ direction. Following Ref.~\cite{Dulat:2018vuy}, we find that we can write
\begin{equation}
\label{eq:azim}
(x_i s_{jk} - x_j s_{ik})^2 = \frac{4 x_i x_j x_k}{\omxk} s_{ij} s_{ijk} \cos^2 \phi_{ij,kl},
\end{equation}
such that $W_{ij}$ has the form,
\begin{equation}
    W_{ij} = \frac{4 x_i x_j x_k}{\omxk} s_{ij} s_{ijk} \left( \cos^2 \phi_{ij,kl} - \frac{1}{2 \ome} \right).
\end{equation}

\section{Results \label{sec:results} }

In this section, we summarise our results for the triple collinear splitting functions.  In each case, we find that the remainder $R_{abc \to P}$ can be expressed in terms of a single trace (rather than three in general). The $\alpha_1$ term from Eq.~\eqref{eq:abasis} is always some combination of two auxiliary functions and they are a feature of the $\alpha$ basis: 
\begin{eqnarray}
\label{eq:A0def}
    A_0(x,y) &=& 1 - \frac{(1-x)}{(1-y)}, \\
\label{eq:B0def}
    B_0(x,y) &=& 1 + \frac{2x(x-2)}{(1-y)^2} + \frac{4x}{(1-y)}.
\end{eqnarray}


\subsection{Three collinear gluons}

We consider the case where gluons $i,j,k$ are in a particular colour-ordering. In other words, the outer gluons $i$ and $k$ play a different role to the inner gluon $j$. We find that, 
\begin{equation}
\label{eq:Pggg}
\Pggg(i,j,k) = \frac{\Pgg(x_i)}{s_{ijk}}  \frac{\Pgg\left(\frac{x_k}{1-x_i} \right)}{s_{jk}} + \frac{\Pgg(x_k)}{s_{ijk}}  \frac{\Pgg\left(\frac{x_i}{1-x_k} \right)}{s_{ij}} +\frac{1}{s_{ijk}^2} \Rggg(i,j,k) ,
\end{equation}
We define $\Rgggsub (i,j,k)$ as the contribution where a ``hard'' gluon $i$ radiates a potentially soft $j,k$ pair. This exposes the manifest $i,k$ symmetry between the outer gluons. We find that,
\begin{equation}
\Rggg (i,j,k) = \Rgggsub (i,j,k) + \Rgggsub (k,j,i),
\end{equation}
where
\begin{eqnarray}
\label{eq:Rggg}
\Rgggsub(i,j,k) &=&  
\frac{2 \ome W_{jk}}{(1-x_i)^2 s_{jk}^2} 
+ \frac{4 \ome x_k}{(1-x_i)^2} \frac{ \Tr{i}{j}{k}{\ell} }{ s_{jk} } \nonumber \\
&& + \fa_0(x_i,x_j,x_k) 
 +  \fa(x_i,x_j,x_k) \frac{s_{ijk}\Tr{i}{j}{k}{\ell}}{s_{ij} s_{jk} } ,
\end{eqnarray}
and
\begin{eqnarray}
\label{eq:fa0}
\fa_0(x_i,x_j,x_k) &=& \ome B_0(x_k,x_i)  ,\\
\label{eq:fa}
\fa(x_i,x_j,x_k) &=& -\frac{x_k \Pgg(x_k)}{x_j(1-x_i)} 
- \frac{\Pgg(x_j)}{x_k} + \frac{2}{x_j(1-x_k)}  -1 - \frac{1}{(1-x_i)(1-x_k)} .
\end{eqnarray}

We note that $\fa$ contains poles in $x_i$, $x_j$ and $x_k$. Therefore, we write $\fa$ in a manner that exposes the residue of these poles, in terms of two-particle splitting functions. Eqs.~\eqref{eq:Pggg}--\eqref{eq:fa} are equivalent to Eq.~(5.4) in Ref.~\cite{campbell} up to a normalisation of a factor of 4. 

As expected, there are no internal single collinear limits (i.e. relating to any of the single collinear limits ($s_{ij} \to 0$, $s_{jk} \to 0$ or $s_{jk} \to 0$) present in Eq.~\eqref{eq:Rggg}.  All of the internal single collinear limits are contained in the iterated contribution.
However, there are possible external and internal singularities when 
\begin{itemize}
\item[(i)] gluon $I$ (for $I \in \{i,j,k\}$) is collinear with the spectator particle $\ell$, indicated when there is one singular power of $x_I$,
\item[(ii)] gluon $I$ is soft, indicated when there are two singular factors in the set $\{s_{IJ}, s_{IK}, x_I\}$.
\end{itemize}
These collinear and/or soft singularities can be present in the $\Pgg \times \Pgg$ contribution and/or in the remainders.  Within the remainders, they are produced entirely by the final term in Eq. (\ref{eq:Rggg}) when,
\begin{equation}
    \fa(x_i,x_j,x_k) \propto \frac{1}{x_I}.
\end{equation}
Given that 
\begin{equation}
\Tr{i}{j}{k}{\ell} = x_k s_{ij} - x_j s_{ik} + x_i s_{jk},
\end{equation}
there are two types of contribution in the remainders. Let us consider the two cases in turn:
\begin{itemize}
    \item $I = k$ (or $I = i$)
\begin{equation}
    \frac{1}{x_k} \frac{s_{ijk}\Tr{i}{j}{k}{\ell}}{s_{ij} s_{jk} } \longrightarrow 
      \frac{x_i}{x_k} \frac{s_{ijk}}{s_{ij}} 
    - \frac{x_j}{x_k} \frac{s_{ijk} s_{ik}}{s_{ij} s_{jk}},
\end{equation}
    
    \item $I = j$
\begin{equation}
\label{eq:Jsoft}
    \frac{1}{x_j} \frac{s_{ijk}\Tr{i}{j}{k}{\ell}}{s_{ij} s_{jk} } \longrightarrow 
    \frac{x_k}{x_j} \frac{s_{ijk}}{s_{jk}} 
    + \frac{x_i}{x_j} \frac{s_{ijk}}{s_{ij}} 
    - \frac{s_{ijk}s_{ik}}{s_{ij}s_{jk}}.
\end{equation}

\end{itemize}

\begin{itemize}

\item[(i)] Gluon $I$ is collinear with the spectator particle $\ell$ - external collinearity.

Let us first consider the external limits where the particle with small momentum fraction is collinear to the spectator particle, $s_{I\ell} = x_I \to 0$. These singular structures are tabulated in Table~\ref{table:ggg}.

The $x_i \to 0$ and $x_k \to 0$ contributions are shown in the first and third rows of Table~\ref{table:ggg}.  These limits are related by the $i \leftrightarrow k$ symmetry, so let us focus on the $x_k \to 0$ limit in the third row.  All contributions are proportional to $\Pgg(x_i)$. They originate in the iterated two-particle splitting and the second term of Eq.~\eqref{eq:fa}.  Note that by construction, there are no contributions from $\Rgggsub(k,j,i)$. 

In the $x_j \to 0$ limit, there are contributions from the iterated two-particle splitting and the double unresolved $\Rgggsub$ splitting.  However, these contributions cancel and the $\Pggg$ splitting function does not exhibit a singularity in this limit. This is as expected, since gluon $j$ is only colour-connected to gluons $i$ and $k$.

\item[(ii)] Gluon $I$ is soft.

The soft $I$ limit is obtained when those in the set $\{s_{IJ}, s_{IK}, x_I\}$ are small and there are two inverse powers of them. The external soft contributions of the form $1/s_{IJ}/x_I$ can be read off from Table~\ref{table:ggg}. However there are also internal soft $j$ contributions coming from the third term in Eq.~\eqref{eq:Jsoft}.
  
When gluon $k$ is soft, we recover the expected limit describing collinear gluons $i$ and $j$ with the soft gluon $k$ radiated between the colour-connected partners $j$ and $\ell$,
\begin{equation}
    \Pggg(i,j,k) \stackrel{k~{\rm soft}}{\longrightarrow} \frac{2 x_j}{s_{jk}x_k} \frac{1}{s_{ij}}\Pgg(x_i).
\end{equation}
This limit comes entirely from the iterated two-particle splitting.
The soft $i$ limit is obtained by $k \leftrightarrow i$ symmetry.  

In the soft $j$ limit, the $1/x_j/s_{ij}$ and $1/x_j/s_{jk}$ terms cancel between the $\Pgg \times \Pgg$ and $\Rgggsub$ contributions, such that
\begin{eqnarray}
    \frac{1}{s_{ijk}^2}\Rgggsub(i,j,k) &
    \stackrel{j~{\rm soft}}{\longrightarrow}& \left(
    -\frac{x_k}{x_j s_{jk}} 
    -\frac{x_i}{x_j s_{ij}} 
    +\frac{s_{ik}}{s_{ij}s_{jk}}
    \right)
    \frac{2}{s_{ik}} \PggS(x_k),\\    
    \Pggg(i,j,k) &
    \stackrel{j~{\rm soft}}{\longrightarrow}& \frac{2 s_{ik}}{s_{ij}s_{jk}}\frac{1}{s_{ik}}\Pgg(x_k).
\end{eqnarray}
This is precisely as expected for the emission of a soft gluon between the hard (and collinear) radiators $i$ and $k$.
\end{itemize}

The limit where both $j$ and $k$ are soft encodes $x_j \to 0$, $x_k \to 0$ and therefore $x_i \to 1$.  There are two types of contribution. First, there are iterated double soft singularities in $\Pgg \times \Pgg$,
\begin{equation} 
\label{eq:DsoftinPxPggg}
\frac{\Pgg(x_i)}{s_{ijk}}  \frac{\Pgg\left(\frac{x_k}{1-x_i} \right)}{s_{jk}} + \frac{\Pgg(x_k)}{s_{ijk}}  \frac{\Pgg\left(\frac{x_i}{1-x_k} \right)}{s_{ij}}
\stackrel{j,k~{\rm soft}}{\longrightarrow}
\frac{2}{\omxi s_{ijk} s_{jk}} \Pgg\left(\frac{x_k}{1-x_i}\right) + \frac{4}{x_j x_k s_{ijk} s_{ij}}.
\end{equation}
Second, there are double soft contributions in $\Rgggsub(i,j,k)$,
\begin{eqnarray}
\label{eq:DsoftinRggg}
\frac{1}{s_{ijk}^2}
\Rgggsub(i,j,k) \stackrel{j,k~{\rm soft}}{\longrightarrow}
 \frac{2 \ome W_{jk}}{(1-x_i)^2 s_{jk}^2 s_{ijk}^2}
 - \left( 
 \frac{2}{x_k (1-x_i)} + \frac{4}{x_j\omxi}
 \right) \frac{\Tr{i}{j}{k}{\ell}}{s_{ij} s_{jk} s_{ijk}}.
\end{eqnarray}
The second term in Eq.~\eqref{eq:DsoftinRggg} is produced by $\fa(x_i,x_j,x_k)$.

The double soft singularities, when gluons $i,j$ are soft, are obtained by the $i \leftrightarrow k$ interchange in Eqs.~\eqref{eq:DsoftinPxPggg} and \eqref{eq:DsoftinRggg}. 

Finally, there are also double soft singularities when gluons $i,k$ are soft, however, because they are not colour-adjacent, they only appear in the $\Pgg \times \Pgg$ contributions as a product of two eikonal factors.

We note that projecting the splitting function onto the $\alpha$-basis of Eq.~\eqref{eq:abasis} forces a link between the trace-like structures and the $B_0$ terms that appear in $\fa_0$, which is evident in the $x_i \to 1$ limit. This corresponds to the $x_j \to 0$, $x_k \to 0$ limit because the three momentum fractions sum to unity. We see that the second and third terms of Eq.~\eqref{eq:Rggg} are separately singular in this limit, 
\begin{eqnarray}
\frac{4\ome x_k}{\omxi^2}\frac{\Tr{i}{j}{k}{\ell}}{s_{jk}} &\to& \phantom{-}\frac{4\ome x_k}{\omxi^2},\nonumber \\
\fa_0(x_i,x_j,x_k) =  \ome B_0(x_k,x_i) &\to& - \frac{4\ome x_k}{\omxi^2},
\end{eqnarray}
and that the singular behaviour cancels when the terms are combined.
The link between the trace-like structures and the $B_0$ terms (including $A_0$ terms in generality) is a feature of the $\alpha$-basis of Eq.~\eqref{eq:abasis} and is repeated in all of the triple collinear splitting functions. 

\begin{landscape}
\begin{table}[p]
\centering
\begin{center}
\begin{tabular}{|c|| c |c| c||c|} 
 \hline
 $\nameggg$ & 
 $\frac{\Pgg(x_i)}{s_{ijk}}  \frac{\Pgg\left(\frac{x_k}{1-x_i} \right)}{s_{jk}} + (i \leftrightarrow k) $ 
 & $\frac{1}{s_{ijk}^2} \Rgggsub(i,j,k)$ 
 & $\frac{1}{s_{ijk}^2} \Rgggsub(k,j,i)$ 
 & $\frac{1}{s_{ijk}^2} \Pggg(i,j,k)$ \\  
 \hline\hline
 $x_i \rightarrow 0$ 
 &\makecell[l]{   \\ $+\frac{1}{s_{ij}s_{ijk}} \frac{ x_j}{x_i} \bigg[2 \Pgg(x_k) \bigg] $ \\$+ \frac{1}{s_{jk}s_{ijk}} \frac{1}{x_i} \bigg[2 \Pgg(x_k) \bigg]$}
 & \makecell[c]{  \\ 0 \\} 
 & \makecell[l]{ $+\frac{1}{s_{ij}s_{jk}} \frac{x_j}{x_i} \bigg[ \Pgg(x_k) \bigg] $ \\ + $\frac{1}{s_{ij}s_{ijk}} \frac{x_j}{x_i} \bigg[ -\Pgg(x_k) \bigg] $ \\ +$\frac{1}{s_{jk}s_{ijk}} \frac{1}{x_i} \bigg[ -\Pgg(x_k)\bigg]$} 
 & \makecell[l]{ $+\frac{1}{s_{ij}s_{jk}} \frac{ x_j}{x_i} \bigg[\Pgg(x_k) \bigg]$ \\ $+ \frac{1}{s_{ij}s_{ijk}} \frac{ x_j}{x_i} \bigg[\Pgg(x_k) \bigg]$ \\$+ \frac{1}{s_{jk}s_{ijk}} \frac{ 1}{x_i} \bigg[ \Pgg(x_k) \bigg]$} \\ 
 \hline
 $x_j \rightarrow 0$ 
 & \makecell[l]{  \\  $+ \frac{1}{s_{ij}s_{ijk}} \frac{ x_i}{x_j} \bigg[ 2 \Pgg(x_k) \bigg]  $  \\ $+ \frac{1}{s_{jk}s_{ijk}} \frac{ x_k}{x_j} \bigg[ 2 \Pgg(x_k) \bigg]$} 
 & \makecell[l]{  \\ $+ \frac{1}{s_{ij}s_{ijk}} \frac{ x_i}{x_j} \bigg[ -2 \PggS(x_k) \bigg]  $ \\$+ \frac{1}{s_{jk}s_{ijk}} \frac{ x_k}{x_j} \bigg[ -2 \PggS(x_k) \bigg]$} 
 & \makecell[l]{  \\ $+ \frac{1}{s_{ij}s_{ijk}} \frac{ x_i}{x_j} \bigg[ -2 \PggS(1-x_k) \bigg]  $ \\$+ \frac{1}{s_{jk}s_{ijk}} \frac{ x_k}{x_j} \bigg[ -2\PggS(1-x_k) \bigg]$} 
 & \makecell[c]{  \\ 0 \\} \\
 \hline
 $x_k \rightarrow 0$ 
 & \makecell[l]{   \\ $+\frac{1}{s_{ij}s_{ijk}} \frac{ 1}{x_k} \bigg[2 \Pgg(x_i) \bigg] $ \\$+ \frac{1}{s_{jk}s_{ijk}} \frac{x_j}{x_k} \bigg[2 \Pgg(x_i) \bigg]$ } 
 & \makecell[l]{  $+\frac{1}{s_{ij}s_{jk}} \frac{x_j}{x_k} \bigg[ \Pgg(x_i) \bigg] $ \\ + $\frac{1}{s_{ij}s_{ijk}} \frac{1}{x_k} \bigg[ -\Pgg(x_i)  \bigg] $ \\ +$\frac{1}{s_{jk}s_{ijk}} \frac{x_j}{x_k} \bigg[ -\Pgg(x_i) \bigg]$} 
 & \makecell[c]{  \\ 0 \\}  
 & \makecell[l]{ $+\frac{1}{s_{ij}s_{jk}} \frac{ x_j}{x_k} \bigg[\Pgg(x_i) \bigg]$ \\ $+ \frac{1}{s_{ij}s_{ijk}} \frac{1}{x_k} \bigg[\Pgg(x_i) \bigg]$ \\$+ \frac{1}{s_{jk}s_{ijk}} \frac{ x_j}{x_k} \bigg[ \Pgg(x_i) \bigg]$} \\
 \hline
\end{tabular}
\end{center}
\caption{Singular behaviour of the $\Pggg$ triple collinear splitting function in the limit where individual momentum fractions are small. The contributions from the iterated two-particle splittings are shown in column 2, while the contributions from the two permutations of $\Rgggsub$ are shown in column 3 and 4 and the contributions for the entire splitting function $\Pggg$ is shown in column 5. Each row shows the singular limit for a different momentum fraction tending to zero. The vertical displacement within each cell is organised by $\{s_{ij},s_{jk},s_{ik},s_{ijk}\}$. }
\label{table:ggg}
\end{table}
\end{landscape}

\subsection{Two gluons with a collinear quark or antiquark}

There are two distinct splitting functions representing the clustering of two gluons and a quark which depend on whether or not the gluons are symmetrised over.

\vspace{3mm}\noindent (a) 
In the case where gluon $j$ is colour-connected to quark $i$ and gluon $k$, we find that,
\begin{eqnarray}
\label{eq:Pqgg}
\Pqgg(i,j,k) &=& 
\frac{\Pqg(x_k)}{s_{ijk}} \frac{\Pqg\left(\frac{x_j}{1-x_k}\right)}{s_{ij}}
+
\frac{\Pqg(1-x_i)}{s_{ijk}}   \frac{\Pgg\left( \frac{x_j}{1-x_i} \right)}{s_{jk}} 
\nonumber \\
&&+\frac{1}{s_{ijk}^2} \Rqgg(i,j,k),
\end{eqnarray}
where
\begin{eqnarray}
\label{eq:Rqgg}
\Rqgg (i,j,k) &=& 
\frac{2 \ome}{(1-x_i)^2}  \frac{W_{jk}}{s_{jk}^2} 
+ \frac{4 \ome x_k }{(1-x_i)^{2}} \frac{\Tr{i}{j}{k}{\ell}}{s_{jk}} + \frac{\ome^2}{(1-x_k)}  \frac{\Tr{i}{j}{k}{\ell}}{s_{ij}}
\nonumber \\
&&
+ \fb_0 (x_i,x_j,x_k) 
+  \fb(x_i,x_j,x_k) \frac{s_{ijk} \Tr{i}{j}{k}{\ell}}{s_{ij}s_{jk}}, 
\end{eqnarray}
and
\begin{eqnarray}
\label{eq:fb0}
\fb_0 (x_i,x_j,x_k) &=& 
\ome \left(B_0(x_k,x_i) -1 + \ome A_0(x_i,x_k)\right),\\
\label{eq:fb}
\fb(x_i,x_j,x_k) &=&- \frac{x_j \Pqg(x_j)}{x_k (1-x_i)} - \frac{2 x_k \Pqg(x_k)}{x_j (1-x_i)} + \frac{4}{(1-x_i)} - 3 \ome.
\end{eqnarray}
Eqs.~\eqref{eq:Pqgg}--\eqref{eq:fb} are equivalent to Eq.~(5.5) in Ref.~\cite{campbell} up to a normalisation of a factor of 4. By charge conjugation, we also have,
\begin{equation}
P_{ \bar q g g  \to \bar q} (i,j,k) = \Pqgg (i,j,k).
\end{equation}

We observe that $\fb$ contains inverse powers of $x_j$ and $x_k$. The 
behaviour of the $\Pqgg$ triple collinear splitting function in the limit where individual momentum fractions are small is tabulated in Table~\ref{table:qgg}.
We see that there is no singular behaviour as $x_i \to 0$.  This reflects the fact that there is no singularity when the quark and spectator momentum are collinear and that there is no soft quark singularity.  When $x_j \to 0$, we see that there are contributions from both the strongly-ordered contribution and from $\Rqgg$ which cancel in the full $\Pqgg$ splitting function, 
\begin{equation}
    \Pqgg(i,j,k) \stackrel{x_j \to 0}{\longrightarrow} 0.
\end{equation}
When $x_k \to 0$, we see that the contributions from the strongly-ordered contribution and from $\Rqgg$ do not cancel in full $\Pqgg$ splitting function.

\begin{table}[t]
\centering
\begin{center}
\begin{tabular}{|c|| c |c|| c|} 
 \hline
 $\nameqgg$ & \makecell[l]{
 $\phantom{+} \frac{\Pqg(x_k)}{s_{ijk}} \frac{\Pqg\left(\frac{x_j}{1-x_k}\right)}{s_{ij}}$
 \\
 + $\frac{\Pqg(1-x_i)}{s_{ijk}}   
 \frac{\Pgg\left( \frac{x_j}{1-x_i} \right)}{s_{jk}}$ 
 } 
 & $\frac{1}{s_{ijk}^2} \Rqgg(i,j,k)$ 
 & $\frac{1}{s_{ijk}^2} \Pqgg(i,j,k)$ \\  
 \hline\hline
 $x_i \rightarrow 0$ 
 &\makecell[c]{0}
 & \makecell[c]{0}
 & \makecell[c]{0} \\ 
 \hline
 $x_j \rightarrow 0$ 
 & \makecell[l]{  \\  $+ \frac{1}{s_{ij}s_{ijk}} \frac{ x_i}{x_j} \bigg[ 2 \Pqg(x_k) \bigg]  $  \\ $+ \frac{1}{s_{jk}s_{ijk}} \frac{ x_k}{x_j} \bigg[ 2 \Pqg(x_k) \bigg]$} 
 & \makecell[l]{  \\  $+ \frac{1}{s_{ij}s_{ijk}} \frac{ x_i}{x_j} \bigg[ -2 \Pqg(x_k) \bigg]  $  \\ $+ \frac{1}{s_{jk}s_{ijk}} \frac{ x_k}{x_j} \bigg[ -2 \Pqg(x_k) \bigg]$} 
 & \makecell[c]{  \\ 0 \\} \\
 \hline
 $x_k \rightarrow 0$ 
 & \makecell[l]{   \\ $+\frac{1}{s_{ij}s_{ijk}} \frac{ 1}{x_k} \bigg[2 \Pqg(x_j) \bigg] $ \\$+ \frac{1}{s_{jk}s_{ijk}} \frac{x_j}{x_k} \bigg[2 \Pqg(x_j) \bigg]$ } 
 & \makecell[l]{ $+ \frac{1}{s_{ij}s_{jk}} \frac{x_j}{x_k} \bigg[ \Pqg(x_j) \bigg] $ \\ $+\frac{1}{s_{ij}s_{ijk}} \frac{ 1}{x_k} \bigg[- \Pqg(x_j) \bigg] $ \\$+ \frac{1}{s_{jk}s_{ijk}} \frac{x_j}{x_k} \bigg[-\Pqg(x_j) \bigg]$} 
 & \makecell[l]{$+ \frac{1}{s_{ij}s_{jk}} \frac{x_j}{x_k} \bigg[ \Pqg(x_j) \bigg] $ \\ $+\frac{1}{s_{ij}s_{ijk}} \frac{ 1}{x_k} \bigg[\Pqg(x_j) \bigg] $   \\$+ \frac{1}{s_{jk}s_{ijk}} \frac{x_j}{x_k} \bigg[\Pqg(x_j) \bigg]$} \\
 \hline
\end{tabular}
\end{center}
\caption{Singular behaviour of the $\Pqgg$ triple collinear splitting function in the limit where individual momentum fractions are small. }
\label{table:qgg}
\end{table}

In the soft $k$ limit, only the strongly-ordered term contributes and we recover the expected limit describing collinear partons $i$ and $j$ with the soft gluon $k$ radiated between the colour-connected partners $j$ and $\ell$,
\begin{equation}
    \Pqgg(i,j,k) \stackrel{k~{\rm soft}}{\longrightarrow} \frac{2x_j}{s_{jk}x_k} \frac{1}{s_{ij}}\Pqg(x_j).
\end{equation}

However, in the soft $j$ limit the $1/x_j/s_{ij}$ and $1/x_j/s_{jk}$ terms cancel between the $P \times P$ and $\Rqgg$ contributions, such that
\begin{eqnarray}
    \frac{1}{s_{ijk}^2} \Rqgg(i,j,k) &
    \stackrel{j~{\rm soft}}{\longrightarrow}& \left(
    -\frac{2x_i}{x_j s_{ij}} 
    -\frac{2x_k}{x_j s_{jk}} 
    +\frac{2s_{ik}}{s_{ij}s_{jk}}
    \right)
    \frac{1}{s_{ik}}\Pqg(x_k),\\    
    \Pqgg(i,j,k) &
    \stackrel{j~{\rm soft}}{\longrightarrow}& \frac{2 s_{ik}}{s_{ij}s_{jk}}\frac{1}{s_{ik}}\Pqg(x_k).
\end{eqnarray}
This is precisely as expected for the emission of a soft gluon between the hard (and collinear) radiators $i$ and $k$.

As in the three gluon splitting function, there are double soft singularities when gluons $j,k$ are soft. These are contained iteratively in the $P \times P$ contributions and in $\Rqgg (i,j,k)$, and are identical to Eqs.~(\ref{eq:DsoftinPxPggg},\ref{eq:DsoftinRggg}),
\begin{eqnarray} 
\label{eq:DsoftinPxPqgg}
\frac{\Pqg(1-x_i)}{s_{ijk}}  \frac{\Pgg\left(\frac{x_j}{1-x_i} \right)}{s_{jk}} && + \frac{\Pqg(x_k)}{s_{ijk}}  \frac{\Pqg\left(\frac{x_j}{1-x_k} \right)}{s_{ij}}\nonumber \\
&&\stackrel{j,k~{\rm soft}}{\longrightarrow}
\frac{2}{\omxi s_{ijk} s_{jk}} \Pgg\left(\frac{x_j}{1-x_i}\right) + \frac{4}{x_j x_k s_{ijk} s_{ij}},\\
\label{eq:DsoftinRqgg}
\frac{1}{s_{ijk}^2}
\Rqgg(i,j,k) &&\stackrel{j,k~{\rm soft}}{\longrightarrow}
 \frac{2 \ome W_{jk}}{(1-x_i)^2 s_{jk}^2 s_{ijk}^2}
 - \left( 
\frac{2}{x_k (1-x_i)} + \frac{4}{x_j\omxi}
 \right) \frac{\Tr{i}{j}{k}{\ell}}{s_{ij} s_{jk} s_{ijk}}. \nonumber \\
\end{eqnarray}
There are no other double soft singularities.

As noted earlier, Eq.~\eqref{eq:Rqgg} also appears to have spurious singularities in both the $x_i \to 1$ and $x_k \to 1$ limits. As in the three gluon splitting function, the singular $x_i \to 1$ behaviour present in the second term in Eq.~\eqref{eq:Rqgg} cancels against the $B_0(x_k,x_i)$ term in $\fb_0$.  The singularity as $x_k \to 1$ in the third term cancels against a similar singularity produced by the $A_0(x_i,x_k)$ term in $\fb_0$.

\vspace{3mm}\noindent (b)  
In the case where the gluons are abelianised ($\tilde{g}$) or two photons are collinear to the quark, then the splitting function is symmetric under the exchange of the two bosons ($j,k$).  We find,
\begin{eqnarray}
\label{eq:Pqpp}
\Pqpp(i,j,k) &=& 
		\frac{\Pqg(x_k)}{s_{ijk}}  
		\frac{\Pqg\left(\frac{x_j}{1-x_k}\right)}{s_{ij}} 
	+	\frac{\Pqg(x_j)}{s_{ijk}}  
		\frac{\Pqg\left(\frac{x_k}{1-x_j}\right)}{s_{ik}} \nonumber \\
&&		+ \frac{1}{s_{ijk}^2} \Rqpp (i,j,k),
\end{eqnarray}
where
\begin{eqnarray}
\label{eq:Rqpp}
\Rqpp (i,j,k) &=&
  -  \frac{\ome^2}{(1-x_k)} \frac{\Tr{j}{i}{k}{\ell}}{s_{ij}} 
  -  \frac{\ome^2}{(1-x_j)} \frac{\Tr{j}{i}{k}{\ell}}{s_{ik}} \nonumber \\
&& +  \fc_0(x_i,x_j,x_k)
		+ \fc(x_i,x_j,x_k) 
		\frac{s_{ijk} \Tr{j}{i}{k}{\ell}}{s_{ij}s_{ik}},
\end{eqnarray}
and
\begin{eqnarray}
\label{eq:fc0}
    \fc_0(x_i,x_j,x_k) &=& \ome\left(2 - \ome A_0(x_j,x_k) -\ome A_0(x_k,x_j)\right), \\
\label{eq:fc}
	\fc(x_i,x_j,x_k) &=& - \frac{x_j \Pqg(x_j)}{x_k (1-x_i)}  - \frac{x_k \Pqg(x_k)}{x_j (1-x_i)} + \frac{4}{(1-x_i)} - 4 \ome + \ome^2.
\end{eqnarray}
Eqs.~\eqref{eq:Pqpp}--\eqref{eq:fc} are equivalent to Eq.~(5.6) in Ref.~\cite{campbell} up to a normalisation of a factor of 4. By charge conjugation, we have 
\begin{equation}
P_{\bar q \gamma \gamma  \to \bar q} (i,j,k) = \Pqpp (i,j,k).
\end{equation}

The 
behaviour of the $\Pqpp$ triple collinear splitting function in the limit where individual momentum fractions are small is tabulated in Table~\ref{table:qpp}. As in the previous case, there is no singular behaviour as $x_i \to 0$ reflecting the fact that there is no singularity when the quark and spectator momentum are collinear and that there is no soft quark singularity.
We also see that there are contributions from both the strongly-ordered contribution and from $\Rqpp$ when $x_j \to 0$ and $x_k \to 0$ that do not cancel in the full $\Pqpp$ splitting function. 
However, only the strongly-ordered term contributes in the soft $j$ or soft $k$ limits, 
\begin{eqnarray}
    \Pqpp(i,j,k) &\stackrel{j~{\rm soft}}{\longrightarrow}& \frac{2x_i}{s_{ij}x_j} \frac{1}{s_{ik}}\Pqg(x_k),\\
   \Pqpp(i,j,k) &\stackrel{k~{\rm soft}}{\longrightarrow}& \frac{2x_i}{s_{ik}x_k} \frac{1}{s_{ij}}\Pqg(x_j).   
\end{eqnarray}
It can be seen that the strongly-ordered terms contribute the full double soft $j,k$ limit (a product of two eikonal factors) and there are no contributions from $\fc (x_i,x_j,x_k)$. There are no other double soft singularities.

\begin{table}[H]
\centering
\begin{center}
\begin{tabular}{|c|| c |c|| c|} 
 \hline
 $\nameqpp$ & 
 \makecell[l]{ $\phantom{+}\frac{\Pqg(x_k)}{s_{ijk}} \frac{\Pqg\left(\frac{x_j}{1-x_k}\right)}{s_{ij}}$
\\ + $ (j \leftrightarrow k) $}
 & $\frac{1}{s_{ijk}^2} \Rqpp(i,j,k)$ 
 & $\frac{1}{s_{ijk}^2} \Pqpp(i,j,k)$ \\  
 \hline\hline
 $x_i \rightarrow 0$ 
 &\makecell[c]{0}
 & \makecell[c]{0}
 & \makecell[c]{0} \\ 
 \hline
 $x_j \rightarrow 0$ 
 & \makecell[l]{  
 $\phantom{\bigg[}$ \\ 
 $+\frac{1}{s_{ij}s_{ijk}} \frac{ x_i}{x_j} \bigg[2 \Pqg(x_k) \bigg] $ \\
 $+ \frac{1}{s_{ik}s_{ijk}} \frac{1}{x_j} \bigg[2 \Pqg(x_k) \bigg]$ } 
 & \makecell[l]{ $+ \frac{1}{s_{ij}s_{ik}} \frac{x_i}{x_j} \bigg[\Pqg(x_k) \bigg]$ \\ 
 $+ \frac{1}{s_{ij}s_{ijk}} \frac{x_i}{x_j} \bigg[-\Pqg(x_k) \bigg]$ \\
 $+ \frac{1}{s_{ik}s_{ijk}} \frac{1}{x_j} \bigg[-\Pqg(x_k) \bigg]$} 
 & \makecell[l]{ $+ \frac{1}{s_{ij}s_{ik}} \frac{x_i}{x_j} \bigg[\Pqg(x_k) \bigg]$ \\ 
 $+ \frac{1}{s_{ij}s_{ijk}} \frac{x_i}{x_j} \bigg[\Pqg(x_k) \bigg]$ \\
 $+ \frac{1}{s_{ik}s_{ijk}} \frac{1}{x_j} \bigg[\Pqg(x_k) \bigg]$} \\
 \hline
 $x_k \rightarrow 0$ 
 & \makecell[l]{   $\phantom{\bigg[}$ \\ 
 $+\frac{1}{s_{ij}s_{ijk}} \frac{ 1}{x_k} \bigg[2 \Pqg(x_j) \bigg] $ \\
 $+ \frac{1}{s_{ik}s_{ijk}} \frac{x_i}{x_k} \bigg[2 \Pqg(x_j) \bigg]$ } 
 & \makecell[l]{ $+ \frac{1}{s_{ij}s_{ik}} \frac{x_i}{x_k} \bigg[\Pqg(x_j) \bigg]$ \\ 
 $+ \frac{1}{s_{ij}s_{ijk}} \frac{1}{x_k} \bigg[-\Pqg(x_j) \bigg]$ \\
 $+ \frac{1}{s_{ik}s_{ijk}} \frac{x_i}{x_k} \bigg[-\Pqg(x_j) \bigg]$} 
 & \makecell[l]{$+ \frac{1}{s_{ij}s_{ik}} \frac{x_i}{x_k} \bigg[\Pqg(x_j) \bigg]$ \\
  $+ \frac{1}{s_{ij}s_{ijk}} \frac{1}{x_k} \bigg[\Pqg(x_j) \bigg]$ \\
  $+ \frac{1}{s_{ik}s_{ijk}} \frac{x_i}{x_k} \bigg[\Pqg(x_j) \bigg]$} \\
 \hline
\end{tabular}
\end{center}
\caption{Singular behaviour of the $\Pqpp$ triple collinear splitting function in the limit where individual momentum fractions are small.  }
\label{table:qpp}
\end{table}

\subsection{Quark-antiquark pair with a collinear gluon}

There are also two distinct splitting functions representing the clustering of a gluon with a quark-antiquark pair into a parent gluon.

\vspace{3mm}\noindent (a)  
When the gluon is colour-connected to the antiquark, we find that, 
\begin{eqnarray}
\label{eq:Pgqbq}
\Pgqbq(i,j,k) &=& 
\frac{\Pqq(x_k)}{s_{ijk}} 
\frac{\Pqg\left(\frac{x_i}{1-x_k}\right)}{s_{ij}} 
+
\frac{\Pgg(x_i)}{s_{ijk}} 
\frac{\Pqq\left(\frac{x_k}{1-x_i}\right)}{s_{jk}}\nonumber \\
&&+ 
\frac{1}{s_{ijk}^2} \Rgqbq(i,j,k),
\end{eqnarray}
where
\begin{eqnarray}
\label{eq:Rgqbq}
\Rgqbq(i,j,k) &= & 
- \frac{2}{(1-x_i)^2} \frac{W_{jk}}{s_{jk}^2} 
- \frac{\ome}{(1-x_k)} \frac{\Tr{i}{j}{k}{\ell}}{s_{ij}} 
- \frac{4 x_k}{(1-x_i)^2} \frac{ \Tr{i}{j}{k}{\ell}}{s_{jk}} \nonumber \\
&&
+ \fd_0 (x_i,x_j,x_k) 
+ \fd (x_i,x_j,x_k) \frac{s_{ijk} \Tr{i}{j}{k}{\ell}}{s_{ij}s_{jk}}, 
\end{eqnarray}
and
\begin{eqnarray}
\label{eq:fd0}
    \fd_0 (x_i,x_j,x_k) &=&- B_0(x_k,x_i) + 1 - \ome A_0(x_i,x_k) 
    , \\
\label{eq:fd}    
    \fd (x_i,x_j,x_k) &=& 
    - \frac{\Pqq (x_k)}{x_i (1-x_i)} + \frac{2}{(1-x_i)} + 1 - 2 x_i + \frac{2( x_j -x_k - 2x_j x_k)}{\ome (1-x_i)}. 
\end{eqnarray}
Eqs.~\eqref{eq:Pgqbq}--\eqref{eq:fd} are equivalent to Eq.~(5.8) in Ref.~\cite{campbell} up to a normalisation of a factor of 4.  By charge conjugation, we have that, 
\begin{equation}
\Pgqbq (i,j,k) = P_{g q \bar q \to g} (i,j,k).
\end{equation}

\begin{table}[H]
\centering
\begin{center}
\begin{tabular}{|c|| c |c|| c|} 
 \hline
 $\namegqbq$ & 
\makecell[l]{$\phantom{+}\frac{\Pqq(x_k)}{s_{ijk}} 
\frac{\Pqg(\frac{x_i}{1-x_k})}{s_{ij}}$ 
\\
+
$\frac{\Pgg(x_i)}{s_{ijk}} 
\frac{\Pqq(\frac{x_k}{1-x_i})}{s_{jk}}$}
 & $\frac{1}{s_{ijk}^2} \Rgqbq(i,j,k)$ 
 & $\frac{1}{s_{ijk}^2} \Pgqbq(i,j,k)$ \\  
 \hline\hline
 $x_i \rightarrow 0$ 
 &\makecell[l]{  \phantom{\bigg[} \\ $+\frac{1}{s_{ij}s_{ijk}} \frac{ x_j}{x_i} \bigg[2 \Pqq(x_k) \bigg] $ \\$+ \frac{1}{s_{jk}s_{ijk}} \frac{1}{x_i} \bigg[2 \Pqq(x_k) \bigg]$ }
 & \makecell[l]{ $+ \frac{1}{s_{ij}s_{jk}} \frac{x_j}{x_i} \bigg[ \Pqq(x_k) \bigg]$  \\ $+\frac{1}{s_{ij}s_{ijk}} \frac{ x_j}{x_i} \bigg[-\Pqq(x_k) \bigg] $ \\$+ \frac{1}{s_{jk}s_{ijk}} \frac{1}{x_i} \bigg[-\Pqq(x_k) \bigg]$ }
 & \makecell[l]{$+ \frac{1}{s_{ij}s_{jk}} \frac{x_j}{x_i} \bigg[ \Pqq(x_k) \bigg]$  \\ $+\frac{1}{s_{ij}s_{ijk}} \frac{ x_j}{x_i} \bigg[\Pqq(x_k) \bigg] $ \\$+ \frac{1}{s_{jk}s_{ijk}} \frac{1}{x_i} \bigg[\Pqq(x_k) \bigg]$} \\ 
 \hline
 $x_j \rightarrow 0$ 
 & \makecell[c]{0}
 & \makecell[c]{0}
 & \makecell[c]{0} \\
 \hline
 $x_k \rightarrow 0$ 
 & \makecell[c]{0}
 & \makecell[c]{0}
 & \makecell[c]{0} \\
 \hline
\end{tabular}
\end{center}
\caption{
Singular Behaviour of the $\Pgqbq$ triple collinear splitting function in the limit where individual momentum fractions are small. }
\label{table:gqbq}
\end{table}

The 
behaviour of the $\Pgqbq$  triple collinear splitting function in the limit where individual momentum fractions are small is tabulated in Table~\ref{table:gqbq}. There are no collinear limits between the quark/antiquark and the spectator. There is singular behaviour as $x_i \to 0$.  In the soft $i$ limit only the strongly-ordered term contributes, 
\begin{eqnarray}
    \Rgqbq(i,j,k) &\stackrel{i~{\rm soft}}{\longrightarrow}& 0,\\
    \Pgqbq(i,j,k) &\stackrel{i~{\rm soft}}{\longrightarrow}& \frac{2x_j}{s_{ij}x_i} \frac{1}{s_{jk}}\Pqq(x_k).
\end{eqnarray}

There are double soft singularities when the $q \bar q$ pair are both soft. These are contained iteratively in the $\Pgg \times \Pqq$ contribution and in $\Rgqbq (i,j,k)$,
\begin{eqnarray} 
\label{eq:DsoftinPxPgqbq}
\frac{\Pgg(x_i)}{s_{ijk}}  \frac{\Pqq\left(\frac{x_k}{1-x_i} \right)}{s_{jk}} 
&\stackrel{j,k~{\rm soft}}{\longrightarrow}&
\frac{2}{\omxi s_{ijk} s_{jk}} \Pqq\left(\frac{x_k}{1-x_i}\right),\\
\label{eq:DsoftinRgqbq}
\frac{1}{s_{ijk}^2}
\Rgqbq(i,j,k) &\stackrel{j,k~{\rm soft}}{\longrightarrow}&
 -\frac{2 W_{jk}}{(1-x_i)^2 s_{jk}^2 s_{ijk}^2}.
\end{eqnarray}
We identify the double soft terms in Eq.~\eqref{eq:DsoftinRgqbq} as uniquely double unresolved. There are no other double soft singularities.

\newpage
\vspace{3mm}\noindent (b) 
The QED-like splitting, where the gluon, quark and antiquark form a photon-like colour singlet is given by, 
\begin{eqnarray}
\label{eq:Ppqbq}
\Ppqbq(i,j,k) &=& 
\frac{\Pqq(1-x_k)}{s_{ijk}} 
\frac{\Pqg\left(\frac{x_j}{1-x_k}\right)}{s_{ij}} 
+
\frac{\Pqq(1-x_i)}{s_{ijk}} 
\frac{\Pqg\left(\frac{x_j}{1-x_i}\right)}{s_{jk}} \nonumber \\ 
&&+ 
\frac{1}{s_{ijk}^2} \Rpqbq(i,j,k),
\end{eqnarray}
where
\begin{eqnarray}
\label{eq:Rpqbq}
\Rpqbq(i,j,k) &= & 
 \frac{\ome}{(1-x_k)} \frac{\Tr{i}{j}{k}{\ell}}{s_{ij}} 
+ \frac{\ome}{(1-x_i)} \frac{\Tr{i}{j}{k}{\ell}}{s_{jk}} \nonumber \\
&&
+ \fe_0 (x_i,x_j,x_k) 
+ \fe (x_i,x_j,x_k) \frac{s_{ijk} \Tr{i}{j}{k}{\ell}}{s_{ij}s_{jk}}, 
\end{eqnarray}
and
\begin{eqnarray}
\label{eq:fe0}
    \fe_0 (x_i,x_j,x_k) &=& 
    -2 + \ome A_0(x_i,x_k) + \ome A_0(x_k,x_i)
    , \\
\label{eq:fe}
    \fe (x_i,x_j,x_k) &=& 
    -\frac{\Pqq(x_i)}{x_j}-\frac{\Pqq(x_k)}{x_j} + \frac{2 \epsilon}{\ome} x_j.
\end{eqnarray}
Eqs.~\eqref{eq:Ppqbq}--\eqref{eq:fe} are equivalent to Eq.~(5.10) in Ref.~\cite{campbell} up to a normalisation of a factor of 4. Note that because of charge conjugation this splitting function is symmetric under the exchange of the quark and antiquark $i,k$. 

\begin{table}[t]
\centering
\begin{center}
\begin{tabular}{|c|| c |c|| c|} 
 \hline
$\namepqbq$ & 
  \makecell[l]{
  $\frac{\Pqq(1-x_k)}{s_{ijk}} 
\frac{\Pqg(\frac{x_j}{1-x_k})}{s_{ij}}$ \\
+ $(i\leftrightarrow k)$}
 & $\frac{1}{s_{ijk}^2} \Rpqbq(i,j,k)$ 
 & $\frac{1}{s_{ijk}^2} \Ppqbq(i,j,k)$ \\  
 \hline\hline
 $x_i \rightarrow 0$ 
 & \makecell[c]{0}
 & \makecell[c]{0}
 & \makecell[c]{0} \\
 \hline
 $x_j \rightarrow 0$ 
 &\makecell[l]{ $+\frac{1}{s_{ij}s_{ijk}} \frac{ x_i}{x_j} \bigg[2 \Pqq(x_k) \bigg] $ \\$+ \frac{1}{s_{jk}s_{ijk}} \frac{x_k}{x_j} \bigg[2 \Pqq(x_k) \bigg]$ }
 & \makecell[l]{ $+\frac{1}{s_{ij}s_{ijk}} \frac{ x_i}{x_j} \bigg[-2 \Pqq(x_k) \bigg] $ \\$+ \frac{1}{s_{jk}s_{ijk}} \frac{x_k}{x_j} \bigg[-2 \Pqq(x_k) \bigg]$ }
 & \makecell[c]{0} \\ 
 \hline
 $x_k \rightarrow 0$ 
 & \makecell[c]{0}
 & \makecell[c]{0}
 & \makecell[c]{0} \\
 \hline
\end{tabular}
\end{center}
\caption{Singular behaviour of the $\Ppqbq$ triple collinear splitting function in the limit where individual momentum fractions are small.}
\label{table:pqqb}
\end{table}

The 
behaviour of the $\Ppqbq$  triple collinear splitting function in the limit where individual momentum fractions are small is tabulated in Table~\ref{table:pqqb}. There are no collinear limits between the quark/antiquark and the spectator.  In the $x_j \to 0$ limit, the contributions from the strongly-ordered terms and $\Rpqbq$ cancel.
In the soft $j$ limit only the strongly-ordered term contributes, 
\begin{eqnarray}
    \frac{1}{s_{ijk}^2} \Rpqbq(i,j,k) &\stackrel{j~{\rm soft}}{\longrightarrow}& 
  \left(
  -\frac{2x_i}{x_j s_{ij}}
    -\frac{2x_k}{x_j s_{jk}}
    +\frac{2s_{ik}}{s_{ij}s_{jk}}\right) \frac{1}{s_{ik}}\Pqq(x_k),\\
    \Ppqbq(i,j,k) &\stackrel{j~{\rm soft}}{\longrightarrow}& \frac{2s_{ik}}{s_{ij}s_{jk}} \frac{1}{s_{ik}}\Pqq(x_k).
\end{eqnarray}
There are no double soft singularities.

\subsection{Quark-antiquark pair  with a collinear quark or antiquark}

Finally, we consider the clustering of a quark-antiquark pair ($Q\bar{Q}$) and a quark $q$ to form a parent quark with the same flavour as q. There are two splitting functions, one where the quark flavours are different and one where the quarks have the same flavour.  

\vspace{3mm}\noindent (a) 
For distinct quarks, we have
\begin{equation}
\label{eq:PqQQ}
\PqQQ(i,j,k) =
\frac{\Pqg(1-x_i)}{s_{ijk}} \frac{\Pqq\left( \frac{x_j}{1-x_i} \right)}{s_{jk}} 
+ \frac{1}{s_{ijk}^2} \RqQQ(i,j,k),
\end{equation}
where
\begin{eqnarray}
\label{eq:RqQQ}
\RqQQ(i,j,k) &=&  - \frac{2}{\omxi^2} \frac{W_{jk}}{s_{jk}^2} 
 - \frac{2 x_k}{\omxi^2} \frac{\Tr{i}{j}{k}{\ell}}{ s_{jk}} - \frac{2 x_j}{\omxi^2} \frac{\Tr{i}{k}{j}{\ell}}{ s_{jk}}  \nonumber\\
&& +\ff_0 (x_i,x_j,x_k)  , 
\end{eqnarray}
and
\begin{equation}
\label{eq:ff0}
\ff_0 (x_i,x_j,x_k) = - \frac{1}{2} ( B_0(x_j,x_i) + B_0(x_k,x_i)) + 1 +\epsilon .
\end{equation}

Eqs.~\eqref{eq:PqQQ}--\eqref{eq:ff0} are equivalent to Eq.~(5.12) in Ref.~\cite{campbell}, up to a normalisation of a factor of 4. Note that this splitting function is symmetric under the exchange of the quark and antiquark $j,k$ of the same flavour and by charge conjugation we have that, 
\begin{equation}
\PqQQ (i,j,k) = P_{q Q \bar Q \to q} (i,j,k).
\end{equation}
Note that we have chosen to make this symmetry explicit in the trace structures.
There are no collinear limits between the quark/antiquark and the spectator, and no soft limits.

The double soft singularities when the $Q \bar Q$ pair are both soft are contained iteratively in the $\Pqg \times \Pqq$ contribution and in the $W_{jk}$ term in $\RqQQ (i,j,k)$, and are equivalent to those given in Eqs.~(\ref{eq:DsoftinPxPgqbq},\ref{eq:DsoftinRgqbq}),
\begin{eqnarray} 
\label{eq:DsoftinPxPqQQ}
\frac{\Pqg(1-x_i)}{s_{ijk}}  \frac{\Pqq\left(\frac{x_j}{1-x_i} \right)}{s_{jk}} 
&\stackrel{j,k~{\rm soft}}{\longrightarrow}&
\frac{2}{\omxi s_{ijk} s_{jk}} \Pqq\left(\frac{x_j}{1-x_i}\right),\\
\label{eq:DsoftinRqQQ}
\frac{1}{s_{ijk}^2}
\RqQQ(i,j,k) &\stackrel{j,k~{\rm soft}}{\longrightarrow}&
 -\frac{2 W_{jk}}{(1-x_i)^2 s_{jk}^2 s_{ijk}^2}.
\end{eqnarray}
There are no other double soft singularities.

\vspace{3mm}\noindent (b) 
For identical quarks, we have 
\begin{equation}
\label{eq:Pqqq}
\Pqqq(i,j,k) = 
\frac{1}{s_{ijk}^2} \Rqqq(i,j,k),
\end{equation}
where
\begin{eqnarray}
\label{eq:Rqqq}
\Rqqq(i,j,k) &=& 
-  \frac{2\ome}{\omxi} \frac{\Tr{i}{j}{k}{\ell}}{s_{jk}} 
-  \frac{2\ome}{\omxk} \frac{\Tr{i}{j}{k}{\ell}}{s_{ij}} 
\nonumber \\
&& + \fg_0(x_i,x_j,x_k) + \fg(x_i,x_j,x_k) 
\frac{s_{ijk}\Tr{i}{j}{k}{\ell}}{s_{ij} s_{jk}},
\end{eqnarray}
and
\begin{eqnarray}
\label{eq:fg0}
\fg_0 (x_i,x_j,x_k) &=&  
 -2 \ome (\epsilon + A_0(x_i,x_k) + A_0(x_k,x_i)), \\
\label{eq:fg}
    \fg(x_i,x_j,x_k)  &=& 
    -\frac{2x_j}{\omxi\omxk}
    +\ome \left(\frac{\omxk}{\omxi} + \frac{\omxi}{\omxk}
    +2+\epsilon \right).
\end{eqnarray}
Eqs.~\eqref{eq:Pqqq}--\eqref{eq:fg} are equivalent to Eq.~(5.14) in Ref.~\cite{campbell}, up to a normalisation of a factor of 4. 
Because of charge conjugation,
\begin{equation}
P_{\bar q q \bar q \to \bar q}  (i,j,k) = \Pqqq(i,j,k).
\end{equation}
There are no collinear limits between the quark/antiquark and the spectator, and no soft limits. There are no double soft singularities.

\subsection{N=1 SUSY Identity}
\label{sec:SUSY}

The two particle and three particle splitting functions are related by an $N=1$ supersymmetry (SUSY) identity that relates the mass of the spin-1 gluon to the spin-1/2 gluino. 
The gluino can be identified as a quark in this scenario. At one loop, the two particle cuts of the one-loop self energy are equal, leading to the identity~\cite{Antoniadis:1981zv}
\begin{equation} 
\label{eq:SUSY2}
	\Pgg (x)  + \Pqq (x) = \Pqg (x) + \Pgq (x) ,
\end{equation}
which only holds in the $D=4, \epsilon = 0$ limit. This is because in dimensional regularisation the number of degrees of freedom of the gluon and gluino are not equal and SUSY is broken.
The left-hand side of Eq.~\eqref{eq:SUSY2} are the two particle cuts of the one-loop gluonic self energy - i.e. the splitting functions which split from a gluon,  while on the right are 
those which split from a quark (gluino). 

Similarly the triple collinear splitting functions are related by the three-particle cuts of the two-loop self energies leading to the $N=1$ SUSY identity~\cite{campbell} (where $\epsilon=0$), 
\begin{eqnarray}
\label{eq:SUSY3}
\lefteqn{
	(\Pggg + 2 \Pgqbq + \Ppqbq) (i,j,k)+ (\text{5 perms.}) }\nonumber \\ 
&=&	(2 \Pqgg + \Pqpp + 2 \PqQQ +  \Pqqq ) (i,j,k) + (\text{5 perms.}).
\end{eqnarray}
The strongly-ordered contributions automatically satisfy Eq.~\eqref{eq:SUSY3} through repeated use of Eq.~\eqref{eq:SUSY2}.  The remaining contributions satisfy, 
\begin{eqnarray}
\label{eq:SUSYR3}
\lefteqn{
	(\Rggg + 2 \Rgqbq + \Rpqbq) (i,j,k)+ (\text{5 perms.}) }\nonumber \\ 
&=&	(2 \Rqgg + \Rqpp + 2 \RqQQ +  \Rqqq ) (i,j,k) + (\text{5 perms.}).
\end{eqnarray}
In Eq.~\eqref{eq:SUSYR3}, the terms proportional to each possible kinematic pole structure, $1/s_{ijk}^2$, $1/(s_{ijk}s_{ij})$,  $1/(s_{ij}s_{jk})$, $1/s_{ij}^2$ and cyclic permutations separately cancel. 
This leads to relations amongst the coefficients of the trace-structures, and amongst the functions multiplying $1/s_{ijk}^2$.
Additionally, by analysing terms proportional to $1/(s_{ij}s_{jk})$, the following relationship holds (where $\epsilon =0$):
\begin{equation}
\begin{split}
    & 2 \fa(x_i,x_j,x_k) + 2 \fd(x_i,x_j,x_k) + \fe(x_i,x_j,x_k) + (i \leftrightarrow k) \\
    & =  2 \fb(x_i,x_j,x_k) + \fc(x_j,x_i,x_k) + \fg(x_i,x_j,x_k) + (i \leftrightarrow k) . \\
\end{split}
\end{equation}

\section{Summary} \label{sec:summary}

In this paper, we have rewritten the triple collinear splitting functions $P_{abc \to P}$ in a way that exposes the single and double unresolved limits.  In particular, we have isolated the strongly-ordered iterated contributions as products of the usual spin-averaged two-particle splitting functions (generically $P \times P$) and a remainder function $R_{abc\to P} (i,j,k)$ that is finite when any pair of $\{i,j,k\}$ are collinear.  We considered spin-averaged splitting functions, and paid particular attention to the azimuthal correlations produced when an intermediate gluon splits into two particles.  This configuration is intimately linked to double soft singularities.

To help with the discussion of the unresolved limits, we introduced the notion of internal and external singularities.   

Internal singularities are only associated with small invariants in the set $\{ s_{ij}, s_{ik}, s_{jk}, s_{ijk}\}$ and correspond to single collinear, single soft or triple collinear $i,j,k$ contributions. By construction, 
\begin{itemize}
    \item[-] {\bf Internal} single collinear singularities between a pair of $\{i,j,k\}$ lead to a factor of $1/s_{ij}$ and are captured by the iterated two-particle splitting contributions ($P \times P$). We write $R_{abc \to P}$ in a way that makes it visibly finite in each of these single collinear limits.
    \item[-] {\bf Internal} single soft singularities, when the soft particle is colour-connected to the other two collinear particles, produce terms like $s_{ik}/(s_{ij}s_{jk})$ and appear only in $R_{abc \to P}$. 
\end{itemize}

External singularities reference other particles involved in the scattering - for example, the spectator particles used to define the momentum fractions of the three collinear particles.  This includes external single collinear singularities involving one of the collinear particles and a spectator particle, soft radiation where one of the spectator particles is colour-connected to the collinear particle or other external double unresolved singularities. These show up in the following way,
\begin{itemize}
    \item[-] When {\bf external} single collinear singularities like $1/x_i$ are present in the full $P_{abc \to P}$ splitting function, they are all contained in $P \times P$ terms.
    \item[-] When {\bf external} single collinear singularities like $1/x_i$ {\bf do not} appear in the full $P_{abc \to P}$ splitting function, then any terms proportional to $1/x_i$ in $P \times P$ will cancel with analagous terms coming from $R_{abc \to P}$. 
    \item[-] {\bf External} single soft singularities where the soft particle is colour-connected to a spectator particle produce terms like $x_i/(x_j s_{ij})$ and appear only in the iterated  $P \times P$ terms. 
\end{itemize}
In the triple collinear splitting function, there are two inverse powers of the small invariants.  Double collinear (two pairs of collinear particles), soft-collinear, double soft or other triple collinear limits than $i,j,k$, all depend on singularities involving one or more of the momentum fractions and are all therefore external singularities.  
In particular, the double soft limit requires at least one singular factor involving the momentum fractions and is classed as an external singularity. These singularities appear in both the iterated $P \times P$ terms and in $R_{abc \to P}$. Double soft singularities are always the overlap between triple collinear $\{i,j,k\}$ and external triple collinear singularities.

We find it useful to think of a hard radiator particle that emits possibly unresolved radiation, together with a spectator particle. In the case of the three-gluon splitting function, we have further decomposed $\Rggg$ into two functions, $\Rgggsub(i,j,k)$ and $\Rgggsub(k,j,i)$.  In $\Rgggsub(i,j,k)$, $i$ can be viewed as a hard radiator emitting unresolved radiation $j$ and $k$.  

Our hope is that decomposing the triple collinear splitting functions will prove useful in developing more efficient infrared subtraction schemes, both at NNLO and by extension to the quadruple collinear case, at N3LO.

\appendix

\section{Two-particle splitting functions} \label{sec:collinear}

For reference, we list the QCD two-particle splitting functions (averaged over the initial particles' polarisations) that are used to build the iterated strongly-ordered limits in the triple collinear decomposition:
\begin{eqnarray} 
\label{eqn:Pqg}
\Pqg(x) &=& \frac{2(1-x)}{x} + \ome x,\\
\label{eqn:Pgg}
\Pgg(x) &=& \frac{2(1-x)}{x} + \frac{2x}{(1-x)} + 2x(1-x),\\
\label{eqn:Pqq}
\Pqq(x) &=& 1 -\frac{2x(1-x)}{\ome},
\end{eqnarray}
where the universal soft terms in $\Pqg$ and $\Pgg$ have been made explicit. 
We also define a sub-splitting function for the $gg \to g$ splitting function,
\begin{equation}
\PggS (x) = \frac{2 (1-x)}{x} + x (1-x),
\end{equation}
such that $\PggS (x) + \PggS (1-x) = \Pgg (x)$.

\acknowledgments

We thank Aude Gehrmann-De Ridder, Thomas Gehrmann and Christian Preuss for useful discussions.
We gratefully acknowledge support from the UK Science and Technology Facilities Council (STFC) through grant ST/T001011/1.

\bibliographystyle{jhep}
\bibliography{bibtry}{}
\end{document}